\documentclass[11pt]{article}

\usepackage{amsfonts}
\usepackage{amsmath}
\usepackage{esint}
\usepackage[margin=1in]{geometry}
\usepackage{graphicx}
\usepackage{epstopdf}
\usepackage{amssymb}
\usepackage{cite}
\usepackage{multicol}
\usepackage{caption}
\usepackage{color}

\newcommand{\ds} {\displaystyle}
\newcommand{\Frac}[2]{\ds \frac{#1}{#2}}

\newcommand{\R}{{\mathbb R}}

\title{\bf \LARGE An approach to periodic, time-varying parameter estimation using nonlinear filtering } 
\author{\Large Andrea Arnold\textsuperscript{1*,2,3} and Alun L. Lloyd\textsuperscript{2,3,4}}
\date{}

\begin{document}
\maketitle

\begin{flushleft}\small
\textbf{1} Department of Mathematical Sciences, Worcester Polytechnic Institute, Worcester, MA, USA
\\
\textbf{2} Department of Mathematics, North Carolina State University, Raleigh, NC, USA
\\
\textbf{3} Center for Quantitative Sciences in Biomedicine, North Carolina State University, Raleigh, NC, USA
\\
\textbf{4} Biomathematics Graduate Program, North Carolina State University, Raleigh, NC, USA
\\
\bigskip

\normalsize

* corresponding author: anarnold@wpi.edu

\end{flushleft}

\bigskip

\captionsetup[figure]{labelfont={bf},name={Figure},labelsep=period}
\captionsetup[table]{labelfont={bf},name={Table},labelsep=period}

\section*{Abstract}
Many systems arising in biological applications are subject to periodic forcing.  In these systems the forcing parameter is not only time-varying but also known to have a periodic structure.  We present an approach to estimating periodic, time-varying parameters that imposes periodic structure by treating the time-varying parameter as a piecewise function with unknown coefficients.  This method allows the resulting parameter estimate more flexibility in shape than prescribing a specific functional form (e.g., sinusoidal) to model its behavior, while still maintaining periodicity.  We employ nonlinear filtering, more specifically, a version of the augmented ensemble Kalman filter (EnKF), to estimate the unknown coefficients comprising the piecewise approximation of the periodic, time-varying parameter.  This allows for straightforward comparison of the proposed method with an EnKF-based parameter tracking algorithm, where periodicity is not guaranteed.  

We demonstrate the effectiveness of the proposed approach on two biological examples: a synthetic example with data generated from the nonlinear FitzHugh-Nagumo system, modeling the excitability of a nerve cell, to estimate the external voltage parameter, and a case study using reported measles incidence data from three locations during the pre-vaccine era to estimate the seasonal transmission parameter.  The formulation of the proposed approach also allows for simultaneous estimation of initial conditions and other static system parameters, such as the reporting probability of measles cases, which is vital for predicting under-reported incidence data.

\bigskip

\noindent \textbf{Keywords:} time-varying parameter estimation; periodic structure; nonlinear filtering; ensemble Kalman filter (EnKF); FitzHugh-Nagumo; measles transmission.

\newpage


\section{Introduction}

Many systems arising in biological applications are subject to periodic forcing, such as the seasonal forcing seen in epidemiological systems \cite{Altizer2006, AronSchwartz1984, GrasslyFraser2006} and the daily forcing in circadian rhythms \cite{Saper2005, Leloup2003, Kronauer1982}.  In these types of systems, the forcing parameter is not only time-varying in nature but is also known to have a periodic structure.  While it is possible to use periodic functions (e.g., sinusoids) to approximate the temporal behavior of such parameters \cite{LondonYorke1973, Dietz1976}, the structural restrictions often do not adequately capture the true time evolution of the parameters.  This illustrates the need for methodology to estimate time-varying parameters that is able to maintain known structural characteristics without imposing restrictive evolution models. 

In this work we present an approach to estimating periodic, time-varying parameters using nonlinear filtering.  In particular, we impose periodicity by treating the time-varying parameter as a piecewise function with unknown coefficients, repeated each period over the course of the time series, and then estimate the coefficients using a version of the ensemble Kalman filter (EnKF).  Formulating the problem in this way allows the resulting time-varying parameter estimate to maintain its periodic structure without imposing any further restrictions to shape.  The proposed approach also permits simultaneous estimation of unknown time-invariant parameters associated with the system, including initial conditions, as may be needed for certain applications.  

Most parameter estimation methodology available in the literature is traditionally aimed at estimating time-invariant (static) parameters.  Classical deterministic techniques for solving the static parameter estimation problem typically rely on least squares optimization routines \cite{Marquardt1963, Johnson1992, DennisSchnabel, Banks2014}.  Bayesian approaches include Markov chain Monte Carlo (MCMC) methods \cite{Andrieu2008, Haario1999, Haario2001, Haario2006} and nonlinear filtering (or sequential Monte Carlo) methods such as particle filters \cite{KaipioSomersalo2005, LiuWest2001, Pitt1999, Ionides2006, Arnold2013} and ensemble Kalman-type filters \cite{Evensen1994, Burgers1998, Evensen2009, Arnold2014}.  

In nonlinear filtering algorithms, static parameters are artificially evolved over time with the aim of converging to a constant value.  If the true parameter values do change slowly over time (i.e., change on a scale slower than the dynamics of the system), then allowing the parameters to evolve with a random walk may capture the drift \cite{Voss2004, Hamilton2013, Matzuka2014}.  Parameter tracking with nonlinear filtering can be used to estimate periodic parameters, as in \cite{Voss2004}.  However, while being the least structurally restrictive method, parameter tracking via random walk evolution does not guarantee that the resulting time series estimate will maintain periodicity or any known structural characteristics inherent to the parameter.  Further, special care must be taken in choosing the variance of the random walk in parameter tracking algorithms in order to avoid filter divergence \cite{Jazwinski1970, Houte1998, Hamill2001, Anderson2001, WhitakerHamill2002, Ng2011, Berry2013, Harlim2010, Gottwald2013} and obtain a useful parameter estimate.

Unlike in parameter tracking algorithms, where periodicity is not imposed, the formulation proposed in this work directly imposes periodic structure throughout the estimation process without forcing the parameter to have a sinusoidal shape.  While the coefficients relating to the piecewise formulation could be estimated using various techniques, nonlinear filtering provides a natural framework to accommodate the sequential nature of the data and time-varying behavior of the parameters we aim to estimate.  Moreover, use of nonlinear filtering to estimate the coefficients allows for straightforward comparison of the proposed approach with the parameter tracking algorithms as described.

The paper is organized as follows.  In Section 2, we review the standard static parameter estimation inverse problem and its solution in the Bayesian statistical framework via nonlinear filtering methods, with particular focus on the augmented EnKF.  After describing our proposed method for estimating periodic, time-varying parameters in Section 3, we demonstrate its effectiveness with two numerical examples in Section 4.  We compare the proposed method to an EnKF-based parameter tracking algorithm, where periodicity is not guaranteed, using synthetic data generated from the FitzHugh-Nagumo system, which describes the spiking dynamics of neurons.  We further demonstrate the proposed method by estimating the seasonal transmission parameter in an epidemic model for the spread of measles.  Results are obtained using time-series data of measles case reports from three locations during the pre-vaccine era, specifically the weekly reported cases in England and Wales (1948-1967) and monthly reported cases in New York City (1945-1964) and Baltimore (1928-1960).  The proposed approach is able to well-capture the periodic, time-varying behavior of the seasonal transmission while simultaneously estimating static parameters representing the initial conditions of the system and the reporting probability of measles cases, which is vital for predicting under-reported incidence data.


\section{ Review: Nonlinear Filtering and the Augmented EnKF }

In setting up the static parameter estimation problem, we assume an ordinary differential equation (ODE) model involving both states $x = x(t)$ and parameters $\theta$, whose values may be uncertain or completely unknown, to describe the system dynamics, i.e.,
\begin{equation}\label{eq:GenODEmodel}
 \Frac{dx}{dt} = f(t,x,\theta), \qquad x(0) = x_0,
\end{equation}
where $x = x(t) \in\R^d$ is the state vector, $\theta\in\R^p$ is the unknown (or poorly known) parameter vector, $f:\R\times\R^d\times\R^p\to\R^d$ is the known model function, and $x_0$ is the possibly unknown (or poorly known) initial value.  We further assume discrete, noisy observations $y_k\in\R^m$, $k=1, 2,..., T_\text{obs}$, of some model states, 
\begin{equation}\label{eq:GenObsModel}
 y_k = g(x(t_k),\theta)+ w_k, \qquad 0<t_1<t_2<\ldots<t_{T_\text{obs}}
\end{equation}
where $g:\R^d\times\R^p\to\R^m$, $m\leq d$, is the known observation function and $w_k$ represents the observation error.  The inverse problem is to estimate the parameter vector $\theta$ and the state vector $x(t)$ at given times from the observations $y_k$.  

We approach the solution to the inverse problem from the Bayesian perspective, where unknowns are treated as random variables.  In particular, we focus on the use of nonlinear filtering algorithms, which provide a natural setting for the time-varying parameter estimation considered in this work.  In this section, we review nonlinear filtering algorithms and the augmented ensemble Kalman filter for combined state and parameter estimation, which we utilize to compute the results in Section 4.


\subsection{Nonlinear Filtering Algorithms}
  
In the Bayesian framework, the model states $x$ and parameters $\theta$ are treated as random variables with probability distributions, and their joint posterior density is assembled using Bayes' theorem
\begin{equation}
\pi(x,\theta\mid y) \propto \pi(y\mid x,\theta)\pi(x,\theta)
\end{equation}
where the likelihood function $\pi(y\mid x,\theta)$ indicates how likely it is that the data $y$ are observed if the state and parameter values were known and the prior distribution $\pi(x,\theta)$ encodes any known information on the states and parameters before taking the data into account.    

Filtering methods rely on the use of discrete-time stochastic equations describing the model states and observations to sequentially update the joint posterior density.  Assume a time discretization $t_j$, $j = 0, 1, \dots, T$, with the observations in \eqref{eq:GenObsModel} occurring possibly in a subset of the discrete time instances.  To avoid double indexing of the time discretization between the propagation steps and observation times, let $y_j = \emptyset$ if there is no observation at $t_j$.  Considering both the model states and observations as Markov processes, we can write an evolution-observation model for the stochastic state and parameter estimation problem using discrete-time Markov models.  The state evolution equation 
\begin{equation}\label{Eq:StateEvolution}
 X_{j+1} = F(X_j,\theta) + V_{j+1}, \quad V_{j+1} \sim \mathcal{N}(0,\mathsf{C}_{j+1}) ,
\end{equation}
where $F$ is a known propagation model and $V_{j+1}$ is an innovation process, allows us to compute the forward time propagation of the state variables $X_j$ given parameters $\theta$.  In this work, $F$ computes the numerical solution to the ODE model \eqref{eq:GenODEmodel} at time $t_{j+1}$.  The observation equation 
\begin{equation}\label{Eq:ObsUpdate}
 Y_{j+1} = G(X_{j+1},\theta) + W_{j+1}, \quad W_{j+1} \sim \mathcal{N}(0, \mathsf{D}_{j+1}) ,
\end{equation}
where $G$ is a known operator and $W_{j+1}$ is the observation noise, is analogous to the observation model \eqref{eq:GenObsModel}.  

Denoting by $D_j$ the accumulated observations up to time $t_j$,
\begin{equation}
D_j = \big\{ y_1,y_2,\dots,y_j \big\},
\end{equation}
the aim of Bayesian filtering is to sequentially update the posterior distribution $\pi(x_j,\theta\mid D_j)$ using a two-step, predictor-corrector-type scheme:
\begin{equation}
\pi(x_j, \theta \mid D_j) \ \longrightarrow \ \pi(x_{j+1}, \theta \mid D_j) \ \longrightarrow \ \pi(x_{j+1}, \theta \mid D_{j+1}) .
\end{equation}
The first step (known as the prediction step) uses the state evolution equation \eqref{Eq:StateEvolution} to predict the values of the states at time $t_{j+1}$ without knowledge of the data, while the second step (the analysis step or observation update) uses the observation equation \eqref{Eq:ObsUpdate} to correct that prediction by taking into account the data at $t_{j+1}$.  Note that if there is no data observed at $t_{j+1}$, then $D_{j+1} = D_j$ and the prediction density $\pi(x_{j+1}, \theta \mid D_j)$ is equivalent to the posterior $\pi(x_{j+1}, \theta \mid D_{j+1}) $.  Starting with a prior density $\pi(x_{0}, \theta \mid D_{0})$, $D_0 = \emptyset$, this updating scheme is repeated until the final joint posterior density is obtained when $j = T$.  

\subsection{Augmented Ensemble Kalman Filter}

There are a variety of nonlinear filtering algorithms for state and parameter estimation available in the literature, including particle filters and ensemble Kalman-type filters; see, e.g., \cite{KaipioSomersalo2005, LiuWest2001, Pitt1999, Ionides2006, Arnold2013, Evensen1994, Burgers1998, Evensen2009, Arnold2014}.  In this work we employ the augmented (or joint) EnKF in the style of \cite{Arnold2014}, which accommodates systems of possibly very stiff differential equations.  The algorithm is implemented as follows.  Assume the current density $\pi(x_j,\theta\mid D_j)$ is represented in terms of an ensemble
\begin{equation}\label{Eq:Ensemble}
{\mathcal S}_{j\mid j} = \Big\{ (x_{j\mid j}^1,\theta_{j\mid j}^1),(x_{j\mid j}^2,\theta_{j\mid j}^2),\ldots, (x_{j\mid j}^{N_\text{ens}},\theta_{j\mid j}^{N_\text{ens}})\Big\}
\end{equation}
where each of the ${N_\text{ens}}$ ensemble members comprises a pair of model states $x_{j\mid j}$ and parameters $\theta_{j\mid j}$ at time $t_j$.  

In the prediction step, the states at time $t_{j+1}$ are predicted using the state evolution equation \eqref{Eq:StateEvolution} to form a state prediction ensemble, 
\begin{equation}
x_{j+1\mid j}^n = F(x_{j\mid j}^n,\theta_{j\mid j}^n) + v_{j+1}^n, \quad n = 1,2,\dots,N_\text{ens},
\end{equation}
where $v_{j+1}^n \sim \mathcal{N}(0,\mathsf{C}_{j+1})$ represents error in the model prediction.  As in \cite{Arnold2014}, we use linear multistep methods for the time integration and systematically assign the model error covariance $\mathsf{C}_{j+1} = \mathsf{C}_{j+1}^n$ sequentially using higher order method error control.  
The parameter values are not updated during the prediction step, so 
\begin{equation}\label{Eq:StaticParamsPred}
\theta_{j+1\mid j}^n = \theta_{j\mid j}^n, \quad n = 1,2,\dots,N_\text{ens}.
\end{equation}
Prediction ensemble statistics are computed using augmented state and parameter vectors
\begin{equation}
z_{j+1\mid j}^n = \left[ \begin{array}{c} x_{j+1\mid j}^n \\ \theta_{j+1 \mid j}^n \end{array} \right] \in \R^{d+k}, \quad n=1,2,\dots,N_\text{ens},
\end{equation}
where the prediction ensemble mean is given by
\begin{equation}
\overline{z}_{j+1\mid j} = \Frac{1}{N_\text{ens}}\ds\sum_{n=1}^{N_\text{ens}} z_{j+1\mid j}^n
\end{equation}
and the prediction (or prior) covariance matrix is
\begin{equation}
\mathsf{\Gamma}_{j+1\mid j} = \Frac{1}{N_\text{ens}-1}\ds\sum_{n=1}^{N_\text{ens}} (z_{j+1\mid j}^n - \overline{z}_{j+1\mid j})(z_{j+1\mid j}^n - \overline{z}_{j+1\mid j})^\mathsf{T} .
\end{equation}

When an observation $y_{j+1}$ arrives, an artificial observation ensemble is generated around the true observation, such that
\begin{equation}\label{Eq:ArtObsEns}
y_{j+1}^n = y_{j+1} + w_{j+1}^n, \quad n = 1,2,\dots,N_\text{ens}
\end{equation}
where $w_{j+1}^n \sim \mathcal{N}(0, \mathsf{D}_{j+1})$ represents the observation error.  The artificial observation ensemble is compared to the observation model prediction ensemble, 
\begin{equation}
 \widehat{y}_{j+1}^n = g(x_{j+1\mid j}^n, \theta_j^n), \quad n = 1,2,\dots,N_\text{ens}, 
\end{equation}
which is computed using the observation function $g$ defined in \eqref{eq:GenObsModel}.  The augmented posterior ensemble at time $t_{j+1}$ is then computed by
\begin{equation}\label{Eq:AnalysisUpdate}
z_{j+1\mid j+1}^n = z_{j+1\mid j}^n + \mathsf{K}_{j+1}\big(y_{j+1}^n - \widehat{y}_{j+1}^n\big), \quad n = 1, 2, \dots, N_\text{ens}
\end{equation}
where the Kalman gain is defined as
\begin{equation}\label{Eq:KalmanGain}
\mathsf{K}_{j+1} = \mathsf{\Sigma}_{j+1}^{z\hat{y}}\big(\mathsf{\Sigma}_{j+1}^{\hat{y}\hat{y}}+ \mathsf{D}_{j+1} \big)^{-1}
\end{equation}
with $\mathsf{\Sigma}_{j+1}^{z\hat{y}}$ denoting the cross covariance of the augmented state-parameter predictions $z_{j+1\mid j}^n$ and observation predictions $\widehat{y}_{j+1}^n$, $\mathsf{\Sigma}_{j+1}^{\hat{y}\hat{y}}$ the forecast error covariance of the observation prediction ensemble, and $\mathsf{D}_{j+1}$ the observation noise covariance.  This formulation of the Kalman gain straightforwardly allows for nonlinear observations, as opposed to the more familiar formula for linear observation models \cite{Moradkhani2005}.  Use of the artificial observation ensemble \eqref{Eq:ArtObsEns} ensures that the resulting posterior ensemble in \eqref{Eq:AnalysisUpdate} does not have too low a variance \cite{Burgers1998}.  The posterior means and covariances for the states and parameters are then computed using posterior ensemble statistics, and the process repeats.

In the above treatment, the parameters $\theta$ are assumed to be static, i.e., $d\theta / dt = 0$, and are artificially evolved over time.  The parameter values are not changed in the prediction step \eqref{Eq:StaticParamsPred} and are only updated in the analysis step \eqref{Eq:AnalysisUpdate} at each data arrival through use of the cross-correlation between the parameters and model states encoded in the Kalman gain \eqref{Eq:KalmanGain}.  If the parameter values are thought to change slowly over time, a drift can be added to the parameters in the prediction step by modeling the change in the parameter values as a random walk, thereby replacing \eqref{Eq:StaticParamsPred} with 
\begin{equation}\label{Eq:ParamDrift}
\theta^n_{j+1 \mid j} = \theta^n_{j\mid j} + \xi^n_{j+1}
\end{equation}
where $\xi^n_{j+1} \sim \mathcal{N}(0,\mathsf{E}_{j+1})$.  The covariance matrix $\mathsf{E}_{j+1}$ of the drift term must be carefully chosen a priori for each application considered in order to avoid filter divergence and obtain a useful parameter estimate.  Filter divergence is a situation in which the EnKF becomes overconfident in an incorrect estimate and ignores subsequent data.  This can occur either when the ensemble spread becomes too small (classical filter divergence) or too large (catastrophic filter divergence).  For more details, see, e.g., \cite{Jazwinski1970, Houte1998, Hamill2001, Anderson2001, WhitakerHamill2002, Ng2011, Berry2013, Harlim2010, Gottwald2013}.


\section{ Treatment of Periodic, Time-Varying Parameters }

The nonlinear filtering methods and augmented EnKF reviewed in Section 2 are typically used to estimate static parameters, assuming $d\theta / dt = 0$ and artificially evolving the parameter over time to converge to a constant value.  In the case of a time-varying parameter that is changing at a rate slower than the dynamics of the system, parameter tracking can be employed to estimate the change in the parameter over time, given that an appropriate drift covariance is selected to capture the drift \cite{Voss2004, Hamilton2013, Matzuka2014}.  However, parameter tracking does not account for known structural characteristics of the time-varying parameter throughout the estimation process.  In particular, using the augmented EnKF with parameter tracking does not guarantee that the periodic variation of the parameter is fully captured or maintained in the resulting time series estimate.  While it is possible to model periodic parameters using periodic functions such as sinusoids, the structural restrictions often do not adequately capture the true time evolution of the parameters \cite{LondonYorke1973, Dietz1976}.

The goal of this work is to use instead an approach that will retain periodic structure without imposing a sinusoidal shape.  To this end, we propose to model periodic, time-varying parameters as piecewise functions with unknown coefficients, repeated each period over the course of the time series, and use nonlinear filtering to estimate the coefficients.  For ease of illustration, in this work we choose piecewise functions comprising a sequence of constant parameters, with each constant in the sequence representing the average value of the time-varying parameter over a specified time segment.  Generally written, if $\gamma(t)$ denotes the time varying parameter with known period $p$, we let
\begin{equation}\label{Eq:PCparam}
\gamma(t_j) \ = \ \gamma_{\frac{\ell}{p}\text{mod}(t_j,p)} \ = \ \begin{cases} \gamma_1  & \text{if } t_j \in \Big[0, \Frac{p}{\ell}\Big) \\[1em]  \gamma_2  & \text{if } t_j \in \Big[\Frac{p}{\ell}, \Frac{2p}{\ell}\Big) \\[1em] \ \vdots & \ \ \ \ \ \ \vdots \\  \gamma_{\ell}  & \text{if } t_j \in \Big[\Frac{(\ell-1)p}{\ell}, p\Big)  \end{cases}
\end{equation}
where $\ell$ is the number of constants in the sequence.  This interpretation allows us to employ nonlinear filtering (in particular, the augmented EnKF as described) to estimate the individual constant parameters $\gamma_m$, $m = 1, \dots, \ell$, for each time segment, while freeing the shape constraints on the time-varying function.  This approach can be straightforwardly extended to treat $\gamma(t)$ as, e.g., a piecewise linear spline and use filtering to estimate the spline coefficients, in the same spirit of freeing the shape constraints of the resulting time-varying function.  Note that the proposed approach is not restricted to the use of nonlinear filtering methods to estimate the coefficients; e.g., least squares optimization or MCMC-type sampling could be used to estimate the $\gamma_m$ parameters.  However, using nonlinear filtering methods allows for straightforward comparison of the proposed approach with the parameter tracking algorithms that have been utilized in this setting.

To estimate $\gamma(t)$ as formulated in \eqref{Eq:PCparam} using the augmented EnKF, we assume that the current density $\pi\big(x_j,\gamma_j \mid D_j\big)$ at time $t_j$ is represented in terms of a discrete ensemble      
\begin{equation}
{\mathcal S}_{j\mid j} = \Big\{ \big(x_{j\mid j}^n,(\gamma_1)_{j\mid j}^n, \dots, (\gamma_\ell)_{j\mid j}^n\big) \Big\}_{n=1}^{N_\text{ens}}
\end{equation}
of size $N_\text{ens}$ as in \eqref{Eq:Ensemble} and apply the algorithm as outlined in Section 2.2 to sequentially update the $\ell$ unknown constants $\gamma_m$ comprising $\gamma(t)$.  

For comparison with the proposed method, if $\gamma(t)$ changes at a rate slower than the dynamics of the system, it is possible to select an appropriate drift covariance so that the augmented EnKF with parameter tracking is able to capture the change in the parameter value over time.  In this case, the discrete ensemble at time $t_j$ is given by
\begin{equation}
{\mathcal S}_{j\mid j} = \Big\{ (x_{j\mid j}^n, \gamma_{j\mid j}^n) \Big\}_{n=1}^{N_\text{ens}},
\end{equation}
where each $\gamma_{j\mid j}^n$ represents an estimate of $\gamma(t_j)$.  At the prediction step of the filter, the parameter ensemble drifts according to the random walk \eqref{Eq:ParamDrift} with $\xi^n_{j+1} \sim \mathcal{N}(0,\sigma_\xi^2)$ for some prescribed variance $\sigma_\xi^2$.  We will demonstrate that while careful selection of $\sigma_\xi$ allows parameter tracking to capture the time-varying behavior of the parameter, this method does not guarantee that the underlying periodicity of the parameter is maintained in the resulting time series estimate.

We note that in many applications, the parameter vector $\theta$ defined in the inverse problem \eqref{eq:GenODEmodel}--\eqref{eq:GenObsModel} may generally contain a combination of both static and time-varying parameters.  In this case, $\gamma(t)$ can be considered a subset of $\theta$, and both formulations of the augmented EnKF can be straightforwardly extended to incorporate simultaneous estimation of static and time-varying parameters.  This will be further demonstrated in the results.


\section{Results}

In this section we demonstrate the effectiveness of the proposed approach to estimating periodic, time-varying parameters on two biological examples.  We first consider a synthetic example with data generated from the nonlinear FitzHugh-Nagumo system, which models the excitability of a nerve cell, to estimate the external voltage parameter, and we compare the results of the proposed method with the parameter tracking algorithm.  We then perform a case study using reported measles incidence data from three locations during the pre-vaccine era to estimate the seasonal transmission parameter.


\subsection{Synthetic Example: Estimating the External Voltage Parameter in the FitzHugh-Nagumo System}

The FitzHugh-Nagumo system \cite{FitzHugh1961}
\begin{eqnarray}
\Frac{dx_1}{dt} &=& c \Big( x_2 + x_1 - \Frac{x_1^3}{3} + v(t) \Big) \label{Eq:FHN1}  \\
\noalign{\vskip4pt}
\Frac{dx_2}{dt} &=& -\Frac{1}{c} \Big( x_1 - a + b x_2 \Big) \label{Eq:FHN2}
\end{eqnarray}
is commonly used as a simplified version of the Hodgkin-Huxley system \cite{HodgkinHuxley1952} to model the spiking dynamics of neurons.  The state variable $x_1$ represents the measurable membrane potential of the neuron, while $x_2$ represents an unobservable combined effect of different ionic currents.  The parameters $a = 0.7$, $b = 0.8$, and $c = 3$ are assumed to be known and fixed, while the time-varying external voltage $v(t)$ is unknown.  

\begin{figure}[t!]
\centerline{\includegraphics[width=0.8\textwidth]{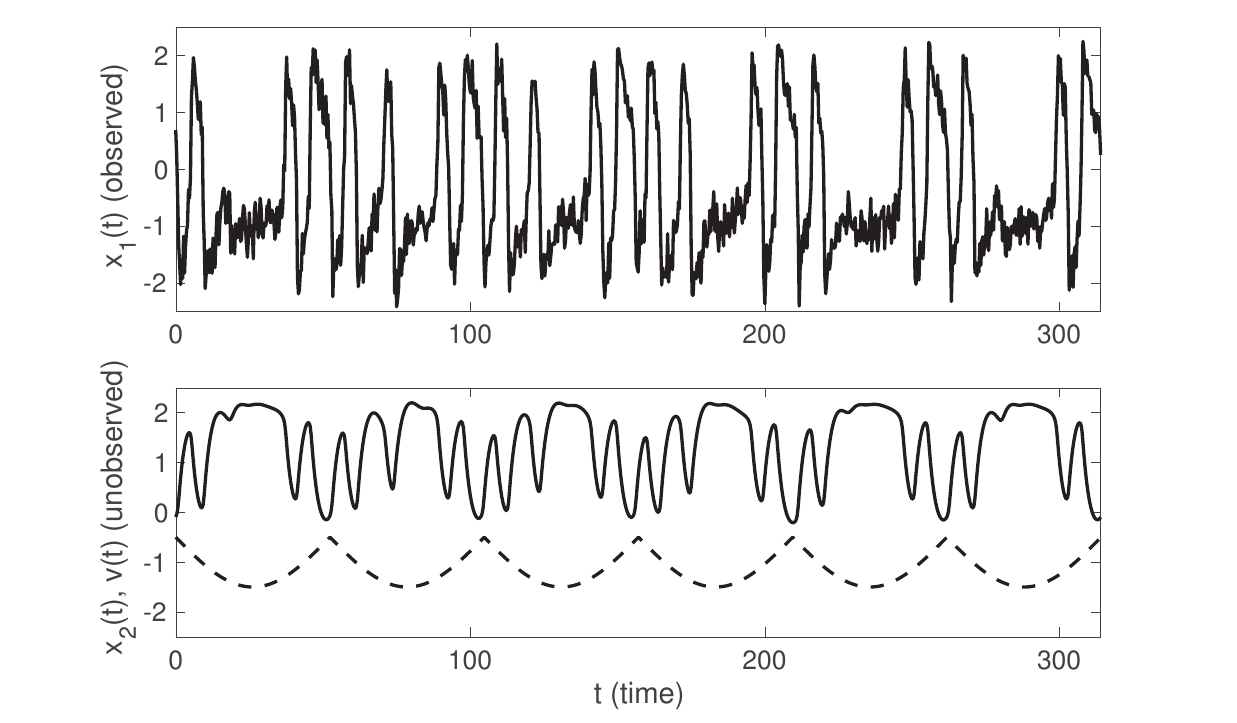}} 
\caption{ {\bf States and voltage parameter of the FitzHugh-Nagumo system. }
Noisy observations of the $x_1$ component (top panel) of the FitzHugh-Nagumo system \eqref{Eq:FHN1}--\eqref{Eq:FHN2}, along with the unobserved state $x_2$ (bottom panel, solid black) and external voltage parameter $v(t)$ (bottom panel, dashed black).  In each panel, the x-axis shows time from $t=0$ to $t=314$ units.}
\label{Fig:SynthDataFHN} 
\end{figure}

We generate synthetic data by letting $v(t)$ be the negative absolute value of a cosine function, up to an additive constant, with frequency $\omega = 0.06$ that varies more slowly than the system dynamics; a similar function is used for $v(t)$ in \cite{Voss2004}.  Measurements of $x_1$ are taken at 943 equidistant time instances over the interval from $t = 0$ to $t = 314$, which covers three full periods of $v(t)$, as shown in Figure \ref{Fig:SynthDataFHN}.  Observations are corrupted with Gaussian noise with zero mean and standard deviation assigned to be 20\% of the standard deviation of the $x_1$ component.  Since $x_2$ is not observed, the observation function 
\begin{equation}
g(x(t_k),\theta) = \mathsf{G} x(t_k)
\end{equation}
is linear, with the projection matrix $\mathsf{G}$ picking out the $x_1$ component of the state vector $x$.  

For the piecewise constant parameter estimation, we treat $v(t)$ as a sequence of 20 constant parameters $v_m$, $m = 1, \dots, 20$, over the course of one period, repeated across all periods as in \eqref{Eq:PCparam}, and use the augmented EnKF with $N_\text{ens} = 200$ ensemble members to estimate $\theta = (v_1, \dots, v_{20})$.  The initial ensemble of parameter values is drawn uniformly from $\mathcal{U}(-2,1)$.  Assuming that the true initial conditions of the system are unknown, we draw the initial state ensemble for $x_1$ uniformly from 0.5 to 1.5 times the first observation point and let the initial state for $x_2$ be zero.  Time integration in the prediction step of the filter is computed using Adams-Moulton methods of orders 2 and 3.  

\begin{figure}[t!]
\centerline{\includegraphics[width=0.9\textwidth]{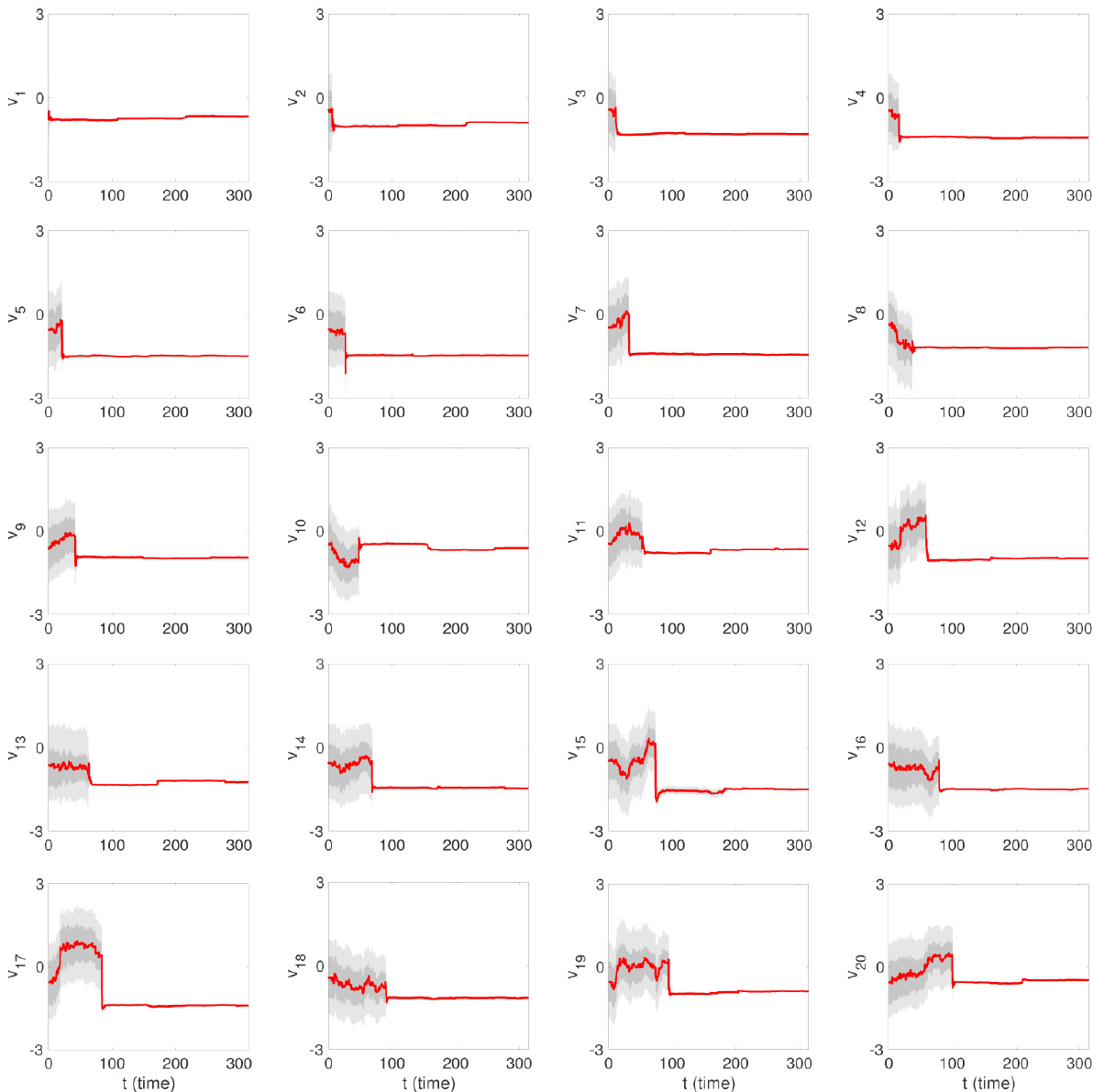}} 
\caption{ {\bf Parameter estimates for the piecewise constant voltage parameter.}
EnKF time series estimates of the constants $v_m$, $m = 1, \dots, 20$, comprising the piecewise constant voltage parameter $v(t)$ in the FitzHugh-Nagumo system \eqref{Eq:FHN1}--\eqref{Eq:FHN2}.  In each panel, the x-axis shows time from $t=0$ to $t=314$ units and the y-axis shows the value of the voltage parameter $v_m$.  The estimated EnKF mean is plotted in solid red, and the 50\% and 90\% credible intervals are plotted in dark and light grey, respectively. } 
\label{Fig:PCparamFHN}
\end{figure}

\begin{figure}[t!]
\centerline{\includegraphics[width=0.9\textwidth]{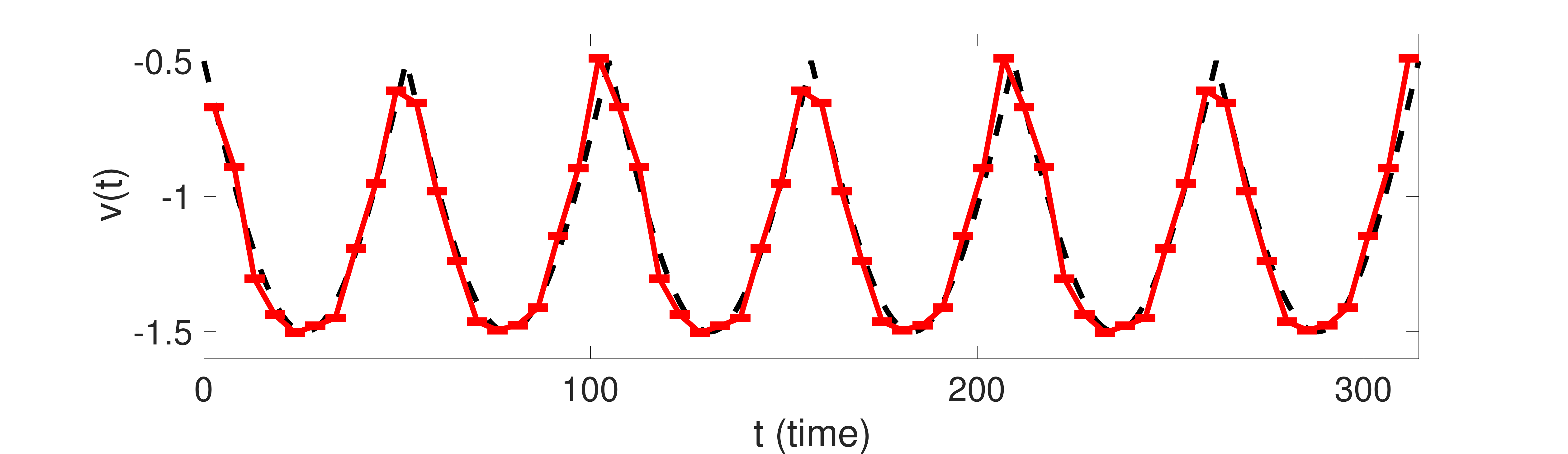}} 
\caption{ {\bf Piecewise estimate of the voltage parameter.}
Posterior estimate of the piecewise constant voltage parameter $v(t)$ in the FitzHugh-Nagumo system \eqref{Eq:FHN1}--\eqref{Eq:FHN2}, repeated over three periods.  The posterior EnKF mean for each $v_m$ is shown in solid red, connected by a linear spline.  The true sinusoidal voltage function used to generate the synthetic data is plotted in dashed black.  }
\label{Fig:PCparamFHN2}  
\end{figure}

Figure \ref{Fig:PCparamFHN} shows the EnKF time series estimates of the constants $v_m$, $m = 1, \dots, 20$, comprising the piecewise constant voltage parameter $v(t)$, and Figure \ref{Fig:PCparamFHN2} shows the resulting estimate of $v(t)$ using the posterior estimates of each $v_m$, repeated over three periods.  Note that the 20 $v_m$ parameters all converge to constant values in a sequential manner over the course of one period with very little uncertainty.  The plot in Figure \ref{Fig:PCparamFHN2} demonstrates that using the posterior mean estimates of the $v_m$ to define $v(t)$ as a piecewise constant function provides a fairly accurate estimate of the underlying sinusoidal voltage function.  Connecting the constant values with a linear spline provides a better visual representation of the estimated $v(t)$ curve.

\begin{figure}[t!]
\centerline{\includegraphics[width=0.9\textwidth]{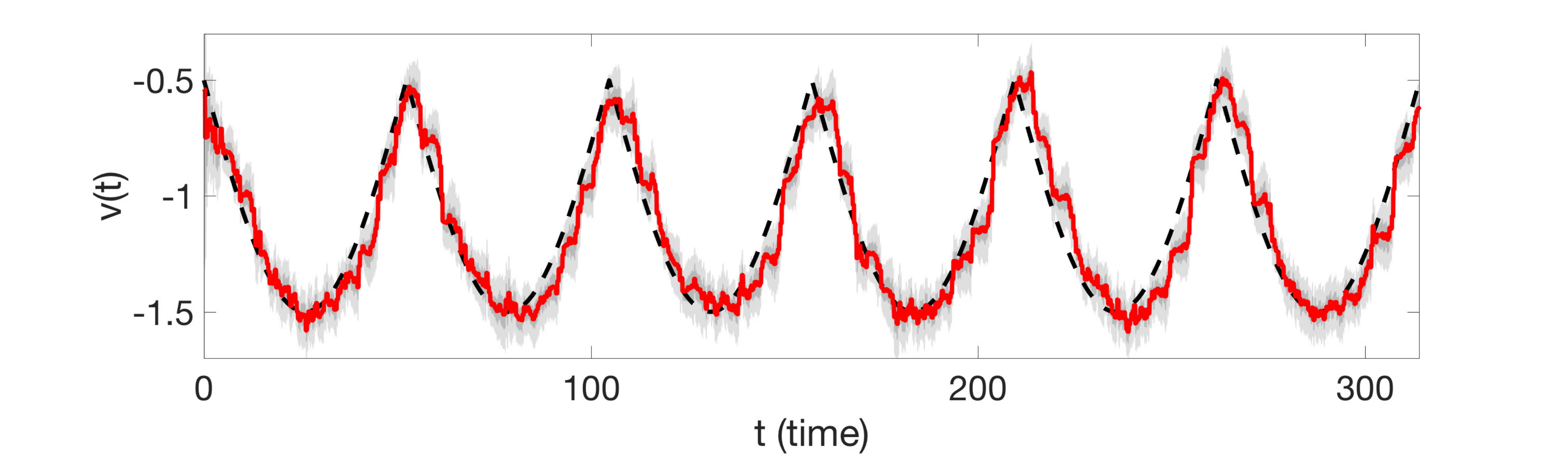}} 
\caption{ {\bf Parameter tracking estimate of the voltage parameter.}
EnKF with parameter tracking estimate of the voltage parameter $v(t)$ in the FitzHugh-Nagumo system \eqref{Eq:FHN1}--\eqref{Eq:FHN2}.  The estimated EnKF mean is plotted in solid red, and the 50\% and 90\% credible intervals are plotted in dark and light grey, respectively.  The true sinusoidal voltage function used to generate the synthetic data is plotted in dashed black.  }
\label{Fig:ParamDriftFHN}  
\end{figure}

For comparison, Figure \ref{Fig:ParamDriftFHN} shows the time series estimate of $v(t)$ using the augmented EnKF with parameter tracking \eqref{Eq:ParamDrift}.  Here $\mathsf{E}_{j+1} = \sigma_\xi^2$ with $\sigma_\xi = 0.01$.  The parameter tracking estimate of $v(t)$ well-captures the overall behavior of the voltage function over time.  However, the estimate is out of phase with and does not maintain the known periodicity of the underlying function.  Although not shown, both methods are able to well-recover the blind system component $x_2$.  The initial conditions of the system could also be estimated, as demonstrated in the next example.

\subsection{Case Study: Estimating the Seasonal Transmission Parameter for Measles Incidence Data}

\begin{figure}[t!]
\centerline{\includegraphics[width=\textwidth]{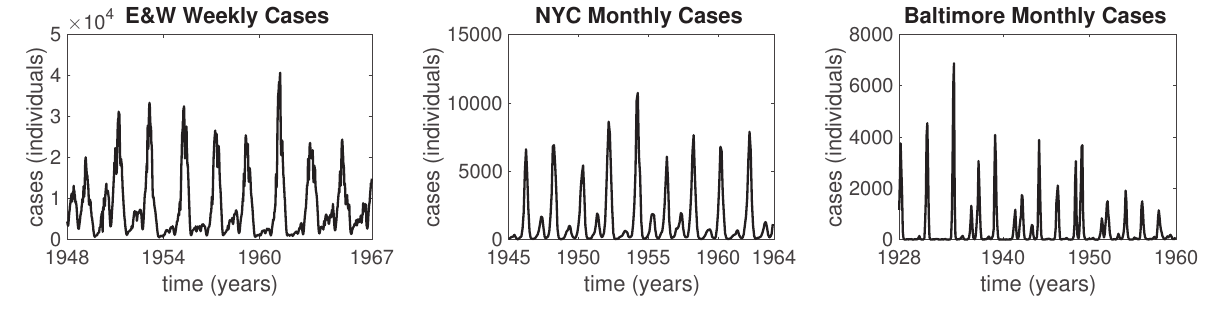}} 
\caption{ {\bf Reported measles cases in three locations during the pre-vaccine era.} 
The weekly reported measles cases in England and Wales (E\&W) from 1948 to 1967 (left panel), monthly reported measles cases in New York City (NYC) from 1945 to 1964 (center panel), and monthly reported measles cases in Baltimore from 1928 to 1960 (right panel).}
\label{Fig:RealData}
\end{figure}

To further demonstrate the effectiveness of the proposed methodology, we perform a case study to estimate the seasonal transmission parameter in an epidemic model for the spread of measles using reported incidence data from three locations during the pre-vaccine era.  The data sets were obtained from an online infectious disease database (http://ms.mcmaster.ca/$\sim$bolker/measdata.html).  In particular, the data comprise the weekly reported measles cases for the aggregate of 60 cities in England and Wales from 1948 to 1967, the monthly reported measles cases in New York City from 1945 to 1964, and the monthly reported measles cases in Baltimore, Maryland, from 1928 to 1960, as shown in Figure \ref{Fig:RealData}.  

Between 1945 and the onset of widespread vaccine usage in the mid 1960's, measles outbreaks in New York City occurred about every two years in the even numbered years.  Estimates show that about 1 in 8 measles cases were reported in New York City during this time \cite{LondonYorke1973}.  Measles outbreaks in Baltimore from 1928 to 1960 were much more sporadic, occurring every two to three years with a less clear periodic pattern.  Approximately 1 in 3 or 4 measles cases were reported in Baltimore during this time \cite{LondonYorke1973}.  The measles outbreaks in England and Wales occurred more regularly, with a nearly annual pattern between 1948 and 1950, then following a bi-annual pattern.  The reporting probability of cases in England and Wales during this time period was found to be relatively high, with greater than 50\% of cases reported \cite{ClarksonFine1985}.  For more details on the data, see \cite{LondonYorke1973, FineClark1982, FinkenstadtGrenfell2000, Bjornstad2002}.  

We use a four compartment Susceptible-Exposed-Infectious-Recovered (SEIR) model to predict the epidemic system dynamics in this application.  For a review of SIR-type models in epidemiology, see, e.g., \cite{AndersonMay1992, Hethcote2000}.  A standard SEIR model comprises the following system of ODEs:
\begin{eqnarray}
\Frac{dS}{dt} &=& m(N-S)-\Frac{\beta(t) SI}{N} \label{eq:dSdt} \label{eq:dSdt} \\
\noalign{\vskip4pt}
\Frac{dE}{dt} &=& \Frac{\beta(t) SI}{N} -(m+a)E\\
\noalign{\vskip4pt}
\Frac{dI}{dt} &=& aE-(m+c)I\\
\noalign{\vskip4pt}
\Frac{dR}{dt} &=& cI-mR \label{eq:dRdt}
\end{eqnarray}
where $m$ is the birth rate of new susceptible individuals and the death rate of individuals in each compartment (assumed here to be equal), $a$ is the per-capita rate at which exposed individuals become infectious, $c$ is the per-capita recovery rate of infectious individuals, and $\beta(t)$ is the unknown seasonal transmission parameter, which varies with time over the course of a year.  Assuming a constant population size $N$, the system \eqref{eq:dSdt}--\eqref{eq:dRdt} can be reduced to
\begin{eqnarray}
\Frac{dS}{dt} &=& m(N-S)-\Frac{\beta(t) SI}{N} \label{eq:dSdt reduced} \\
\noalign{\vskip4pt}
\Frac{dE}{dt} &=& \Frac{\beta(t) SI}{N} -(m+a)E \label{eq:dEdt reduced}  \\
\noalign{\vskip4pt}
\Frac{dI}{dt} &=& aE-(m+c)I \label{eq:dIdt reduced}
\end{eqnarray}
where 
\begin{equation}
R(t) = N - S(t) - E(t) - I(t).
\end{equation}

As described above, the data are the recorded number of measles cases reported over a specified time period (e.g., weekly, monthly).  Since measles cases are known to be under-reported \cite{He2009, Gunning2014}, each observation $y_k = y(t_k)$ is modeled as a reported fraction of the total number of cases accumulated between times $t_{k-1}$ and $t_k$.  The observation model is formulated as in \eqref{eq:GenObsModel} with nonlinear observation function
\begin{equation}\label{Eq:ObsFunc}
g(x(t_k),\theta) = \rho \ d(x(t_k),\theta)
\end{equation}
where $\rho$ is the reporting probability, assumed here to be constant over time, and
\begin{equation}\label{Eq:Cases}
d(x(t_k),\theta) = \ds\int_{t_{k-1}}^{t_k} \Frac{\beta(t)S(t)I(t)}{N} dt
\end{equation}
denotes the total number of cases between times $t_{k-1}$ and $t_k$.  The observation function in \eqref{Eq:ObsFunc} can be written equivalently as
\begin{equation}\label{Eq:ObsFunc2}
g(x(t_k),\theta) = \ds\int_{t_{k-1}}^{t_k} \rho \ \Frac{\beta(t)S(t)I(t)}{N} dt
\end{equation}
where
\begin{equation}
\Frac{dg}{dt} = \rho \ \Frac{\beta(t)S(t)I(t)}{N} 
\end{equation}
gives the rate of change of the cumulative cases between times $t_{k-1}$ and $t_k$.  

We impose an annually-varying, periodic structure on the seasonal transmission parameter by modeling $\beta(t)$ as a sequence of constant parameters $\beta_m$, $m = 1, \dots, 12$, with each $\beta_m$ representing the transmission parameter for a given month, as in \eqref{Eq:PCparam}; i.e., 
\begin{equation}\label{Eq:PCbeta}
\beta(t_j) \ = \ \beta_{\text{mod}(t_j,12)} \ = \ \begin{cases} \beta_1  & \text{if } t_j \in \mbox{January} \\  \beta_2  & \text{if } t_j \in \mbox{February} \\ \ \vdots & \ \ \ \ \ \ \vdots \\  \beta_{12}  & \text{if } t_j \in \mbox{December}  \end{cases} 
\end{equation}
Since we assume that the same seasonal pattern is repeated annually \cite{FinkenstadtGrenfell2000}, the period of the transmission parameter is one year.  
 
In addition to $\beta(t)$, there are several static system parameters that need to be estimated in this application.  While the the birth/death rate $m$, the exposed-to-infectious rate $a$, the recovery rate $c$, and the population size $N$ in \eqref{eq:dSdt reduced}--\eqref{eq:dIdt reduced} can be fairly well estimated from the literature and demographic data, the initial model states $S(0)$, $E(0)$, and $I(0)$ remain uncertain.  Further, while some estimates of the reporting probability $\rho$ can be obtained \cite{He2009, Gunning2014}, this parameter is also uncertain.  Therefore, the unknown parameter vector to be estimated is $\theta = (\beta_1, \dots, \beta_{12}, S(0), E(0), I(0), \rho)\in\R^{16}$.

Synthetic validation was performed on the measles application prior to using the real data; results obtained using synthetic data are provided as supplementary material in the Appendix.  Results using the real data are obtained using a similar procedure to the synthetic examples, incorporating the details specific to each data set in the filter setup.  In each simulation, the regional population sizes are assumed to be constant and are approximated from available demographic data (see http://ms.mcmaster.ca/$\sim$bolker/measdata.html).  In particular, we assume the population size for England and Wales is approximately 40,000,000, New York City is 7,800,000, and Baltimore is 891,080, during the respective time spans over which the measles cases were recorded.  

Prior ensembles of the initial model states are drawn by assuming that the population at each location initially comprises 5\% susceptible individuals, 0.8\% exposed, and 0.2\% infected, then sampling from a uniform distributions between 0.8 and 1.2 times those values.  Prior ensembles for the reporting probabilities are drawn using  available reporting information for each location during the years considered \cite{LondonYorke1973, ClarksonFine1985}.  In particular, for England and Wales the initial distribution of $\rho$ is drawn uniformly between 0.55 and 0.75, for New York City between 0.05 and 0.2, and for Baltimore between 0.2 and 0.4.  

\begin{figure}[t!]
\centerline{\includegraphics[width=\textwidth]{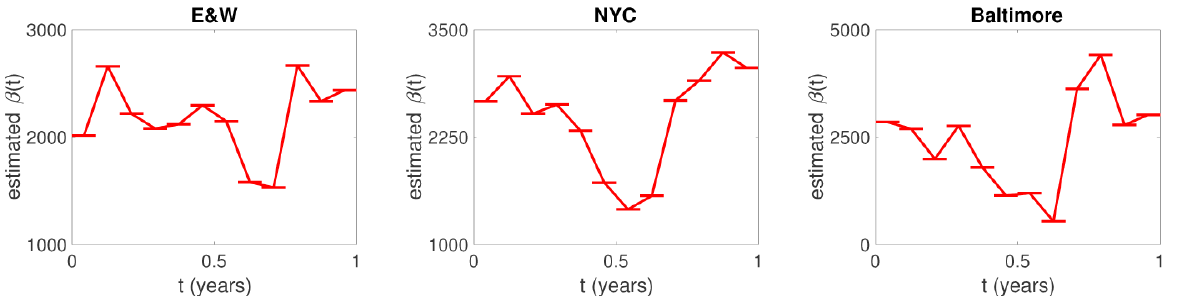}} 
\caption{ {\bf Piecewise estimates of the transmission parameter in each location.} 
Posterior estimates of the piecewise constant transmission parameters $\beta(t)$ corresponding to the reported measles cases in England and Wales (E\&W), New York City (NYC), and Baltimore.  In each figure, the posterior EnKF mean for each constant $\beta_m$, corresponding to each month in a year, is shown in solid red, connected by a linear spline.  Corresponding posterior estimates of the static parameters $S(0)$, $E(0)$, $I(0)$, and $\rho$ for each location are listed in Table \ref{Table:RealDataStaticParams}.}
\label{Fig:RealDataResults}
\end{figure}

Figure \ref{Fig:RealDataResults} shows the resulting piecewise constant estimate of $\beta(t)$ for the three respective locations, using the posterior estimates of each of the 12 estimated $\beta_m$ constants for each data set, connected with linear splines.  Table \ref{Table:RealDataStaticParams} lists the corresponding static parameter estimates for $S(0)$, $E(0)$, $I(0)$, and $\rho$ for each location.  The resulting shapes of the transmission parameters reflect the annual pattern that we might expect, with the lowest value of $\beta(t)$ in each location occurring during the summer months, corresponding to summer holiday break for school-aged children.  

\begin{table}[b!]
\renewcommand{\arraystretch}{2}
\begin{center}
\begin{tabular}{@{} | c | c c c |}
\hline
\ & E\&W & NYC & Baltimore \\
\hline
$S(0)$ & 1,531,000 & 569,300 & 4,1250 \\
$E(0)$ & \  \ 245,000 & \ 91,080 & \ 6,600 \\
$I(0)$ & \ \ \ 61,240 & \ 22,770 & \ 1,650 \\
$\rho$ & \ \ \ 0.7316 & \ 0.1282 & 0.3270 \\
\hline
\end{tabular}
\end{center}
\caption{ \label{Table:RealDataStaticParams}  
{\bf Static parameter estimates for the reported measles cases in each of three locations.} 
Posterior estimates of the static parameters $S(0)$, $E(0)$, $I(0)$, and $\rho$ for the reported measles cases in England and Wales (E\&W), New York City (NYC), and Baltimore, respectively, which correspond to the estimated transmission parameters $\beta(t)$ shown in Figure \ref{Fig:RealDataResults}.  Parameter values are listed with four significant figures.}
\end{table}

As the summer break in the United Kingdom typically begins later than in the United States, the minimum value of the transmission parameter shown in Figure \ref{Fig:RealDataResults} for the England and Wales data occurs a bit later (i.e., in September) than in the US cities (July for New York City, August for Baltimore).  The transmission parameter noticeably rises during the fall months (September, October) when children return to school, before decreasing again during the winter holiday break.  There is also a clear dip in $\beta(t)$ during the spring (March, April), which may coincide with a mid-semester break.  The gradual drop seen in $\beta(t)$ during certain times over the course of the semester may be the result of imperfect mixing; see, e.g., \cite{Bjornstad2002, FinkenstadtGrenfell2000, Grenfell2002}.


\section{Discussion}

In this work we present an approach to estimating periodic, time-varying forcing parameters in nonlinear systems through use of nonlinear filtering methodology.  We demonstrate the effectiveness of the approach using two applications from the life sciences, namely estimating the external voltage parameter in the FitzHugh-Nagumo system for neuronal spiking and the seasonal transmission parameter in an epidemic model for the spread of measles.  By treating the forcing parameters in these applications as piecewise constant functions and using nonlinear filtering methodology to estimate their coefficients, we are able to incorporate the periodic structure of the parameters into the estimation process without prescribing restrictive evolution models.  We are also able to simultaneously estimate time-invariant parameters associated with the system, which for the measles example includes the initial conditions of the model states and the reporting probability of cases for the data considered.  

The approach presented in this work is meant as a step towards better estimating time-varying parameters by incorporating known structural characteristics, in particular, periodicity.  For the applications at hand, treating the forcing parameter as a periodically repeating piecewise constant function is a middle ground approach between assigning a strict functional shape and letting the parameter drift with no guaranteed structure.  Assigning, e.g., a sinusoidal function does not necessarily represent the true behavior of the time-varying parameter of interest for applications with real data.  This can be seen in the results in Figure \ref{Fig:RealDataResults} for the measles data, where the seasonal transmission parameters corresponding to the real data sets analyzed in this work noticeably do not follow strict sinusoidal patterns over the course of a year.  

While the piecewise constant approach requires that $\ell$ constant parameters be estimated, it provides a way to enforce periodicity over the full time series of the data without imposing a restrictive functional form.  The parameter tracking approach, while requiring that fewer parameters be estimated during the filtering process (i.e., one dynamic parameter vs. $\ell$ constants), introduces an additional nuisance parameter (namely, the drift variance $\sigma_\xi^2$) which much be carefully chosen a priori for each application considered in order to avoid filter divergence and obtain a useful estimate.  The parameter tracking approach also does not guarantee the resulting time-series estimate will maintain the periodic structure of the parameter over the full time series, as seen with the FitzHugh-Nagumo external voltage parameter in Figure \ref{Fig:ParamDriftFHN}.  Future work may include combining these two approaches to better incorporate known structural characteristics into a parameter tracking-type filtering algorithm to allow for more freedom in the shape of the resulting time-varying parameter estimate.  

It is indeed possible to use a variety of parameter estimation techniques, such as least squares optimization or MCMC sampling, to estimate the $\ell$ constant coefficients in the piecewise constant approach.  In this work, we highlight the use of nonlinear filtering algorithms, specifically the augmented EnKF, since these methods are amenable to the time-series data often available in life sciences applications and the resulting posterior distributions contain a measure of uncertainty in the parameter estimates.  While not the focus of this work, the posterior parameter distributions can be used for model prediction and uncertainty quantification, which would require additional variational techniques in the deterministic setting \cite{Banks2014,Smith2013}.  Further, using nonlinear filtering to estimate the coefficients in the proposed method allows for straightforward comparison with the parameter tracking algorithm as described.  While we use the augmented EnKF to obtain our results, a variety of other sequential filtering algorithms could also be employed, e.g., a particle filter \cite{KaipioSomersalo2005, LiuWest2001, Pitt1999, Ionides2006, Arnold2013} or a dual filter, either heuristic \cite{Moradkhani2005} or Bayesian-consistent \cite{AitEl2016}.  

As previously noted, the proposed method is not restricted to the use of piecewise constant functions and can be straightforwardly extended to treat the time-varying parameter as, e.g., a piecewise linear spline and use nonlinear filtering to estimate the spline coefficients.  The piecewise constant interpretation used in this work is meant as a simple example of a more general framework for estimating periodic, time-varying parameters in nonlinear systems.  While in this work we consider only one time-varying parameter per system to estimate, additional model parameters, such as the reporting probability $\rho$ in the measles application, could be treated as time-varying, assuming some known structural characteristics and taking into account the added computational cost.


\section*{Acknowledgments}
This work was supported by National Science Foundation grant RTG/DMS-1246991 (Research Training Group in Mathematical Biology at North Carolina State University).



\bigskip
\bigskip



\renewcommand{\thefigure}{A\arabic{figure}}
\renewcommand{\thetable}{A\arabic{table}}

\setcounter{figure}{0}
\setcounter{equation}{0}
\setcounter{table}{0}

\bigskip
\bigskip

\appendix

\section*{Appendix}

\subsection*{Validation of the Method on Synthetic Measles Incidence Data}

Prior to using the real measles data described in the manuscript, we performed validation of the proposed method for estimating periodic, time-varying parameters using synthetic incidence data.  The inverse problem considered is to track the model states $x = (S, E, I)\in\R^3$ of the reduced SEIR model \eqref{eq:dSdt reduced}--\eqref{eq:dIdt reduced} and estimate the time-varying seasonal transmission parameter $\beta(t)$, along with the static model initial conditions $S(0)$, $E(0)$, and $I(0)$ and reporting probability $\rho$, given monthly data on the reported number of measles cases, with the observation function modeled as in \eqref{Eq:ObsFunc2}.  For the problem at hand, we assume that the parameters $N$, $m$, $a$ and $c$ in system \eqref{eq:dSdt reduced}--\eqref{eq:dIdt reduced} are known and fixed, so the unknown parameter vector that we want to estimate is defined as $\theta = (\beta_1, \dots, \beta_{12}, S(0), E(0), I(0), \rho)\in\R^{16}$. 

\begin{figure}[t!]
\centerline{\includegraphics[width=0.75\textwidth]{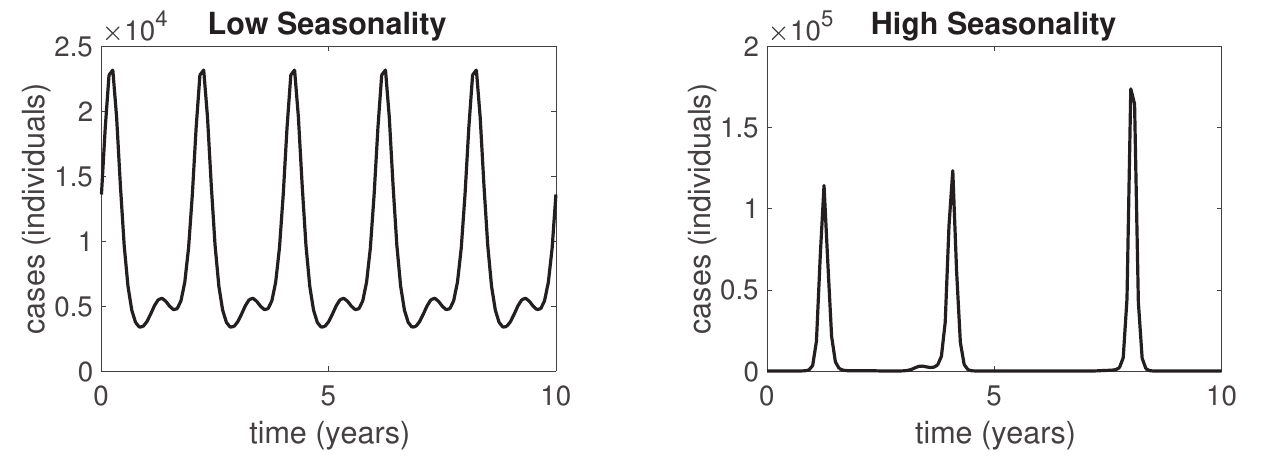}} 
\caption{ {\bf Synthetic measles incidence data.}
Synthetic monthly incidence data generated using the sinusoidal transmission function \eqref{Eq:SinusoidalBeta} with average transmission $b_0 = 1800$ and amplitude $b_1$ chosen to simulate low seasonality ($b_1 = 0.08$) and high seasonality ($b_1 = 0.3$).}
\label{Fig:SynthData}  
\end{figure}

We generate synthetic data using the fixed parameters 
\begin{equation}
N = 9.235\times 10^6, \quad m = 0.02, \quad a = 35.84, \quad c = 100 \stepcounter{equation}\tag{A\theequation}
\end{equation}
and the sinusoidal transmission function 
\begin{equation}\label{Eq:SinusoidalBeta}
\widehat{\beta}(t) = b_0 \Big( 1 + b_1 \cos(2\pi t) \Big)  \stepcounter{equation}\tag{A\theequation}
\end{equation}
with average transmission $b_0 = 1800$ and amplitude of variation $b_1$, which can be viewed as the first term of a Fourier expansion of the underlying seasonal function.  Figure \ref{Fig:SynthData} shows two synthetic data sets, generated to simulate low seasonality (with $b_1 = 0.08$) and high seasonality (with $b_1 = 0.3$), respectively.  In each case, we simulate the model states for 120 years, then discard the transient and record the total number of measles cases per month over 10 years.  We assume that 60\% of cases are reported and that the observations are corrupted by a small amount of Gaussian noise with standard deviation $\sigma = 0.1$.  For demonstration purposes, we show results relating to the low seasonality data; although not shown, similar results are obtained using the high seasonality data.

To validate the proposed approach, in which we treat $\beta(t)$ as a sequence of constant parameters $\beta_m$, $m = 1, \dots, 12$, with each $\beta_m$ representing the transmission parameter for a given month, i.e., 
\begin{equation}\label{Eq:PCbetaA}
\beta(t_j) \ = \ \beta_{\text{mod}(t_j,12)} \ = \ \begin{cases} \beta_1  & \text{if } t_j \in \mbox{January} \\  \beta_2  & \text{if } t_j \in \mbox{February} \\ \ \vdots & \ \ \ \ \ \ \vdots \\  \beta_{12}  & \text{if } t_j \in \mbox{December}  \end{cases}   \stepcounter{equation}\tag{A\theequation}
\end{equation} 
as defined in \eqref{Eq:PCbeta}, we use the augmented EnKF outlined in Section 2.2 with $N_\text{ens} = 250$ ensemble members to estimate the parameters $\theta = (\beta_1, \dots, \beta_{12}, S(0), E(0), I(0), \rho)$ as described.  We draw the initial parameter ensemble, for $n = 1,\dots,N_\text{ens}$, using uniform distributions as follows: $(\beta_m)_{0\mid 0}^n \sim \mathcal{U}(1000,2500)$ for $m = 1, \dots, 12$, $x_{0\mid 0}^n \sim \mathcal{U}(0.25 \widehat{x}_0, 2 \widehat{x}_0)$ where $\widehat{x}_0 = (\widehat{S}(0), \widehat{E}(0), \widehat{I}(0))\in\R^3$ is the initial condition of the states used to generate the data, and $\rho_{0\mid 0}^n \sim \mathcal{U}(0.5, 0.75)$.  Time integration in the prediction step of the filter is computed using Adams-Moulton methods of orders 1 and 2. 

\begin{figure}[b!]
\centerline{\includegraphics[width=\textwidth]{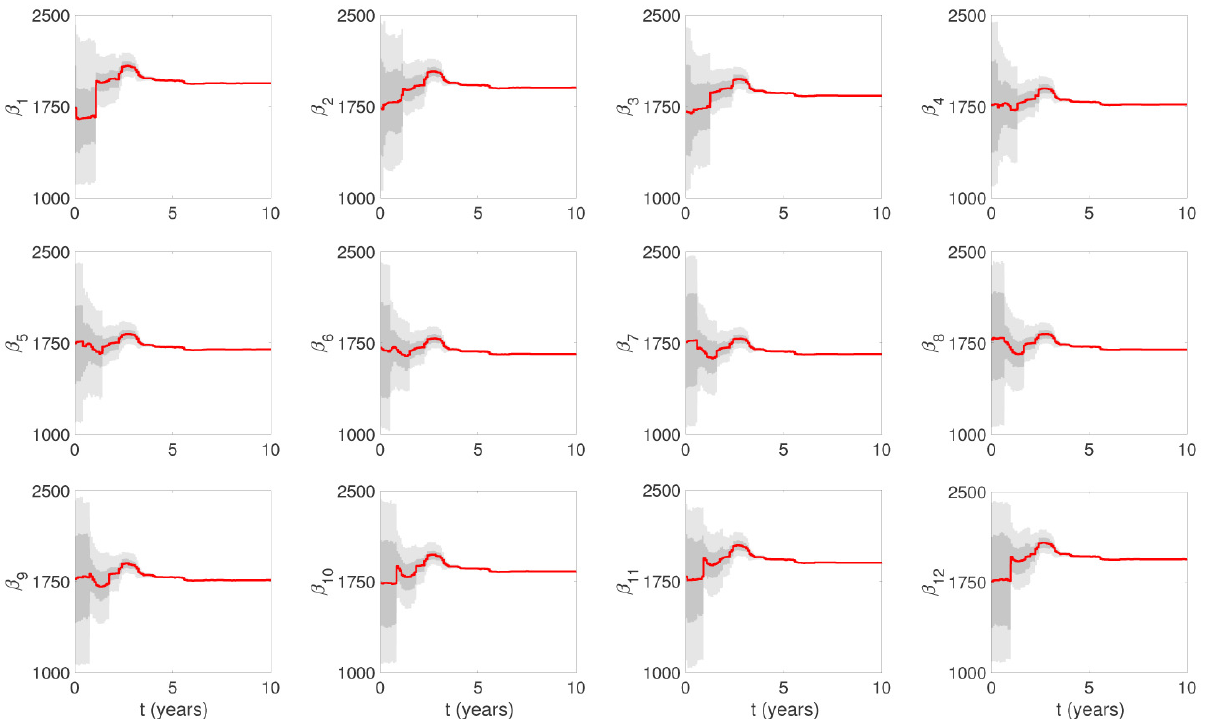}} 
\caption{ {\bf Parameter estimates for the piecewise constant transmission parameter for the low seasonality synthetic data.}
EnKF time series estimates of the constants $\beta_m$, $m = 1, \dots, 12$, comprising the piecewise constant transmission parameter $\beta(t)$ defined in \eqref{Eq:PCbetaA} for synthetic data with low seasonality.  In each panel, the x-axis shows time from 0 to 10 years and the y-axis shows the value of the transmission parameter $\beta_m$.  The estimated EnKF mean is plotted in solid red, and the 50\% and 90\% credible intervals are plotted in dark and light grey, respectively.}
\label{Fig:LowSeasonalityPCbeta}  
\end{figure}

\begin{figure}[!h]
\centerline{\includegraphics[width=\textwidth]{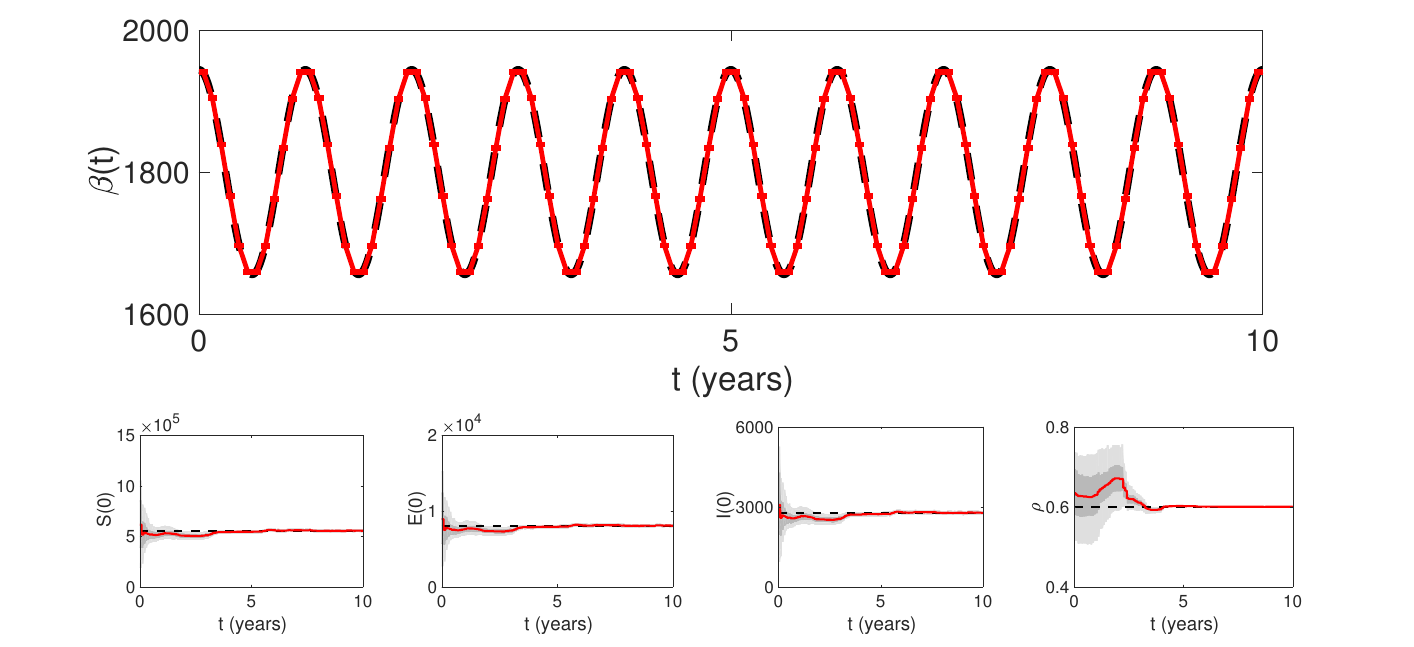}} 
\caption{ {\bf Piecewise estimate of the transmission parameter, with corresponding static parameters, for the low seasonality synthetic data.}
(Top panel) Posterior estimate of the piecewise constant transmission parameter $\beta(t)$ defined in \eqref{Eq:PCbetaA}, repeated over 10 years, for the low seasonality synthetic data.  The posterior EnKF mean for each $\beta_m$, corresponding to each month in a year, is shown in solid red, connected by a linear spline.  The true sinusoidal transmission function $\widehat{\beta}(t)$ in \eqref{Eq:SinusoidalBeta} used to generate the low seasonality synthetic data is plotted in dashed black.  (Bottom panel) EnKF time series estimates of the static parameters $S(0)$, $E(0)$, $I(0)$, and $\rho$, estimated simultaneously along with $\beta_m$, $m = 1, \dots, 12$.  The estimated EnKF mean is plotted in solid red, and the 50\% and 90\% credible intervals are plotted in dark and light grey, respectively.  The true parameter values used to generate the low seasonality synthetic data are plotted in dashed black. }
\label{Fig:LowSeasonalityPCbeta2}  
\end{figure}

Figure \ref{Fig:LowSeasonalityPCbeta} shows the EnKF time series estimates of the constants $\beta_m$, $m = 1, \dots, 12$, comprising the piecewise constant transmission parameter $\beta(t)$ defined in \eqref{Eq:PCbetaA} for the synthetic data with low seasonality.  Figure \ref{Fig:LowSeasonalityPCbeta2} shows the resulting estimate of $\beta(t)$ using the posterior estimates of each $\beta_m$, repeated over 10 years, along with the corresponding time series estimates of the static parameters.  Note that the 12 $\beta_m$ parameters and four additional static parameters all converge to constant values around year 6 with very little uncertainty.  The top panel in Figure \ref{Fig:LowSeasonalityPCbeta2} demonstrates that using the posterior mean estimates of the $\beta_m$ to define the piecewise constant $\beta(t)$ provides a fairly accurate estimate of the true sinusoidal transmission function $\widehat{\beta}(t)$ used to generate the data.  Connecting the piecewise constant values with a linear spline provides a better visual representation of the estimated $\beta(t)$ curve.  

As previously noted, the proposed approach for estimating periodic, time-varying parameters is not restricted to use of the augmented EnKF; various parameter estimation techniques could be utilized to estimate the $\beta_m$ coefficients and additional static parameters.  Table \ref{Table:LScompare} compares the mean estimates of $\beta_m$, $m=1,\dots,12$, and the static parameters $S(0)$, $E(0)$, $I(0)$, and $\rho$ computed using the augmented EnKF formulation (shown in Figures \ref{Fig:LowSeasonalityPCbeta} and \ref{Fig:LowSeasonalityPCbeta2}) with estimates obtained using a least squares optimization routine with a random initial guess.  The optimization was performed utilizing MATLAB's \texttt{fminsearch} function, which employs a Nelder-Mead simplex method for unconstrained nonlinear minimization \cite{NelderMead, DennisJr}, to find the parameter values that minimize the sum of squared errors between the observed case data and the observation model defined in \eqref{Eq:ObsFunc2}.  The initial guess for $\theta$ was chosen similarly to the augmented EnKF prior ensemble, with random draws $(\beta_m)_0 \sim \mathcal{U}(1000,2500)$ for $m = 1, \dots, 12$, $x_0 \sim \mathcal{U}(0.25 \widehat{x}_0, 2 \widehat{x}_0)$, and $\rho_0 \sim \mathcal{U}(0.5, 0.75)$.  The optimizer was set to compute a maximum of 20,000 function evaluations.

The ``true'' values of the $\beta_m$ coefficients listed in Table \ref{Table:LScompare} are the values of the true sinusoidal function $\widehat{\beta}(t)$ in \eqref{Eq:SinusoidalBeta} evaluated at the midpoint of each month over the course of a year (in units of years).  The relative error in each case is computed as
\begin{equation}\label{Eq:RelErr}
\left| \Frac{\theta_\text{true}-\theta_\text{est}}{\theta_\text{true}} \right|  \stepcounter{equation}\tag{A\theequation}
\end{equation}
where $\theta_\text{est}$ is the estimated parameter vector (either by augmented EnKF or least squares) and $\theta_\text{true}$ is the true parameter vector.  The results in Table \ref{Table:LScompare} show that the parameter estimates are fairly similar in both cases, with comparable relative errors.  While not the focus of this work, it is possible to analyze the computational advantages of using nonlinear filtering in this setting, as these algorithms require time integration only from one point to the next, instead of integrating over the full time series of data at each update of the parameter estimate.

\begin{table}[b!]
\renewcommand{\arraystretch}{2}
\begin{center}
\begin{tabular}{@{} | c || c || c | c || c | c |}
\hline
\ & \ \ \ \ True  \ \ \ \ & EnKF Mean & Rel. Error ($\times 10^{-3}$) & Least Squares & Rel. Error ($\times 10^{-3}$)  \\
\hline
$\beta_1$     & \ \ \ 1,939.1  & \ \ \ 1,940.7 &  0.8485   & \ \ \ 1,943.8 & \ \ \  2.4435   \\
$\beta_2$     & \ \ \ 1,901.8   & \ \ \ 1,904.1 & 1.2222    & \ \ \ 1,899.9 & \ \ \  1.0004   \\
$\beta_3$     & \ \ \ 1,837.3   & \ \ \ 1,838.7 & 0.7908    & \ \ \ 1,839.4 & \ \ \  1.1612   \\
$\beta_4$     & \ \ \ 1,762.7  & \ \ \ 1,766.0 &  1.8449   & \ \ \ 1,763.6 & \ \ \   0.4718  \\
$\beta_5$     & \ \ \ 1,698.2  & \ \ \ 1,696.7 & 0.8766    & \ \ \ 1,698.0 & \ \ \  0.1307  \\
$\beta_6$     & \ \ \ 1,660.9  & \ \ \ 1,659.6 &  0.7770   & \ \ \ 1,660.2  & \ \ \   0.4054 \\
$\beta_7$     & \ \ \ 1,660.9  & \ \ \ 1,658.9 &  1.1906   & \ \ \ 1,659.0  & \ \ \   1.1514  \\
$\beta_8$     & \ \ \ 1,698.2  & \ \ \ 1,695.9 &  1.3513   & \ \ \ 1,698.0  & \ \ \  0.0951  \\
$\beta_9$     & \ \ \ 1,762.7  & \ \ \ 1,761.7 &  0.5697   & \ \ \  1,762.5 & \ \ \   0.1579 \\
$\beta_{10}$ & \ \ \ 1,837.3  & \ \ \ 1,834.3 &  1.6356   & \ \ \ 1,837.9  & \ \ \   0.3245 \\
$\beta_{11}$ & \ \ \ 1,901.8  & \ \ \ 1,902.8 &  0.5101   & \ \ \ 1,903.3  & \ \ \   0.7933  \\
$\beta_{12}$ & \ \ \ 1,939.1  & \ \ \ 1,940.6 &  0.7678   & \ \ \ 1,940.1  & \ \ \   0.5110  \\
$S(0)$           & 553,020.0   & 553,730.0   &  1.2727   & 552,370.0     & \ \ \  1.1752   \\
$E(0)$           & \ \ \ 8,042.9  & \ \ \ 8,053.1 &  1.2727   & \ \ \ 8,832.2  &  \ \ 98.1410  \\
$I(0)$            & \ \ \ 2,765.1  & \ \ \ 2,768.7 &  1.2727    & \ \ \ 2,058.7  &  255.4700  \\
$\rho$           & \ \ \ 0.6000  & \ \ \ \ 0.6001 &  0.1230   & \ \ \ \ 0.6002 & \ \ \  0.3096   \\
\hline
\end{tabular}
\end{center}
\caption{ \label{Table:LScompare}  
{\bf Parameter estimates: augmented EnKF vs. least squares optimization.} 
Estimates of the 12 $\beta_m$ parameters comprising the piecewise constant formulation of $\beta(t)$ and the four additional static parameters $S(0)$, $E(0)$, $I(0)$, and $\rho$ obtained using the augmented EnKF and least squares optimization via MATLAB's \texttt{fminsearch}.  The true parameter values are listed along with the posterior mean estimates obtained using the augmented EnKF and the least squares estimates.  The relative errors as defined in \eqref{Eq:RelErr} are given for both methods.  Computed values are listed with five significant figures.}
\end{table}

\begin{figure}[!h]
\centerline{\includegraphics[width=\textwidth]{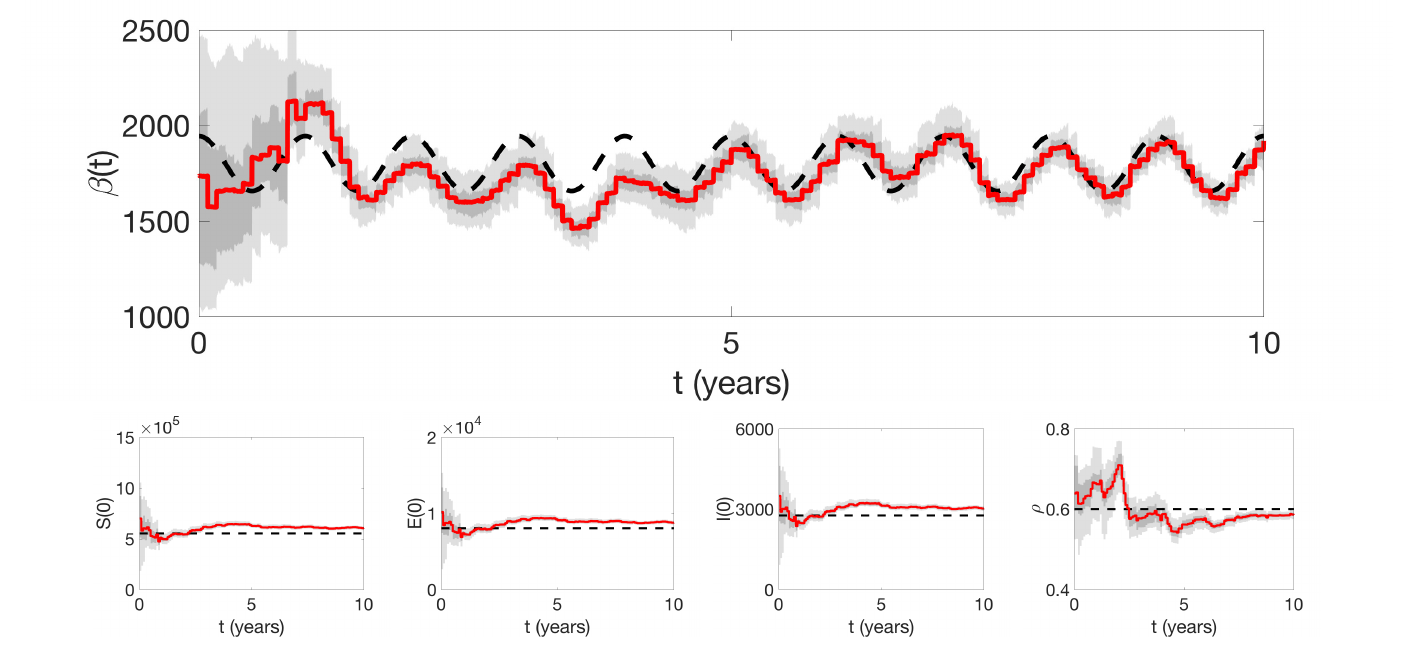}} 
\caption{ {\bf Parameter tracking estimate of the transmission parameter, with corresponding static parameters, for the low seasonality synthetic data.}
(Top panel) EnKF with parameter tracking estimate of the transmission parameter $\beta(t)$ for the low seasonality synthetic data with $\sigma_\xi = 3.8$.  The estimated EnKF mean is plotted in solid red, and the 50\% and 90\% credible intervals are plotted in dark and light grey, respectively.  The true sinusoidal transmission function $\widehat{\beta}(t)$ in \eqref{Eq:SinusoidalBeta} used to generate the low seasonality synthetic data is plotted in dashed black.  (Bottom panel) EnKF time series estimates of the static parameters $S(0)$, $E(0)$, $I(0)$, and $\rho$, estimated simultaneously along with $\beta(t)$.  The estimated EnKF mean is plotted in solid red, and the 50\% and 90\% credible intervals are plotted in dark and light grey, respectively.  The true parameter values used to generate the low seasonality synthetic data are plotted in dashed black. }
\label{Fig:LowSeasonalityParamDrift}  
\end{figure}

For comparison with the proposed approach, Figure \ref{Fig:LowSeasonalityParamDrift} shows the time series estimate of $\beta(t)$ for the low seasonality data using the augmented EnKF with parameter tracking, along with the corresponding time series estimates of the four static parameters.  Here, the covariance of the drift term is set as $\mathsf{E}_{j+1} = \sigma_\xi^2$ with $\sigma_\xi = 3.8$.  After around year 7, the parameter tracking estimate of $\beta(t)$ is able to capture the overall behavior of the transmission.  However, the estimate is out of phase with the true underlying transmission and the resulting shape does not fully capture the annual seasonal variation over the entire time series.  Further, the corresponding posterior estimates of the static parameters $S(0)$, $E(0)$, $I(0)$, and $\rho$ are not as accurate as in the piecewise case.

\begin{figure}[!h]
\centerline{\includegraphics[width=\textwidth]{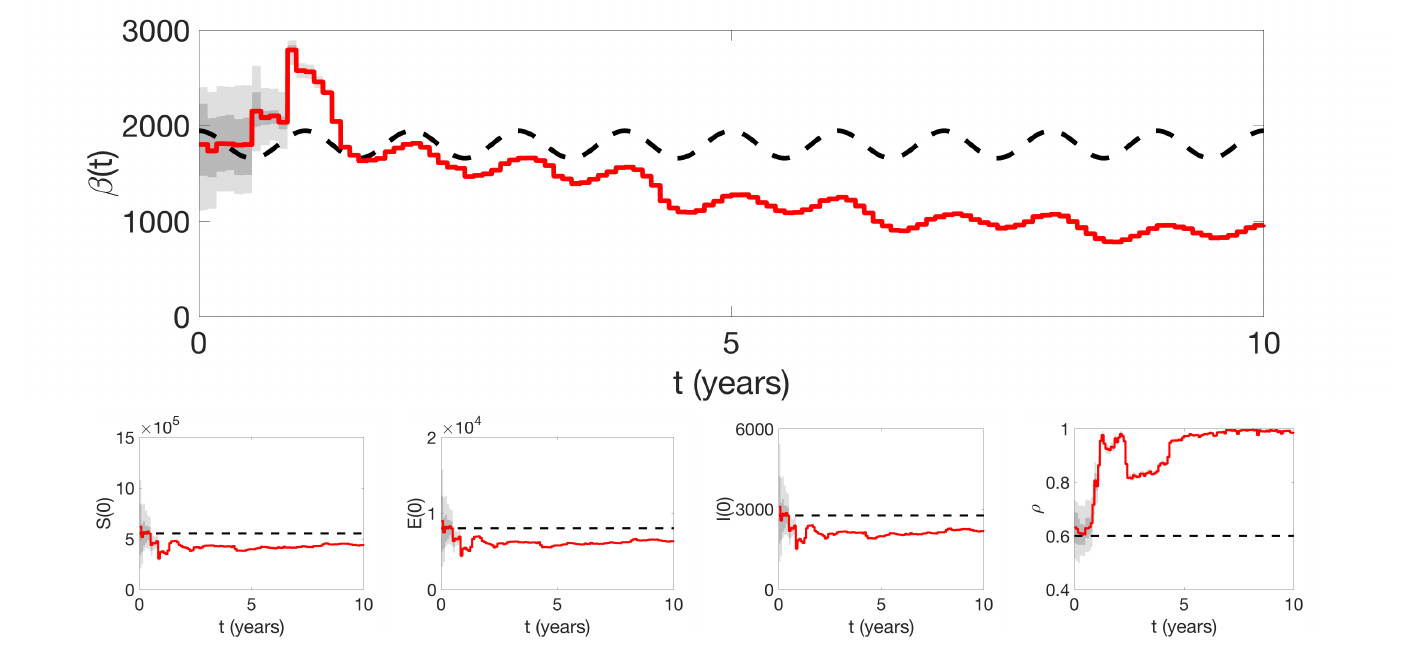}} 
\caption{ {\bf Diverging parameter tracking estimate of the transmission parameter, with corresponding static parameters, for the low seasonality synthetic data.}
(Top panel) EnKF with parameter tracking estimate of the transmission parameter $\beta(t)$ for the low seasonality synthetic data with $\sigma_\xi = 0.5$.  The estimated EnKF mean is plotted in solid red, and the 50\% and 90\% credible intervals are plotted in dark and light grey, respectively.  The true sinusoidal transmission function $\widehat{\beta}(t)$ in \eqref{Eq:SinusoidalBeta} used to generate the low seasonality synthetic data is plotted in dashed black.  (Bottom panel) EnKF time series estimates of the static parameters $S(0)$, $E(0)$, $I(0)$, and $\rho$, estimated simultaneously along with $\beta(t)$.  The estimated EnKF mean is plotted in solid red, and the 50\% and 90\% credible intervals are plotted in dark and light grey, respectively.  The true parameter values used to generate the low seasonality synthetic data are plotted in dashed black. }
\label{Fig:Diverge}  
\end{figure}

As previously noted, the covariance matrix $\mathsf{E}_{j+1}$ of the drift term in the parameter tracking algorithm must be chosen carefully in order to avoid filter divergence and obtain a useful parameter estimate.  The parameter tracking estimate for this example is sensitive to the choice of the drift variance $\sigma_\xi^2$.  To demonstrate this, Figure \ref{Fig:Diverge} shows the resulting parameter tracking estimate of $\beta(t)$ for the low seasonality data with $\sigma_\xi = 0.5$.  Note that in this case the parameter estimate diverges around year 3, which is mirrored in the tracking of the susceptible population $S(t)$ (not shown) and also affects the estimates of the static parameters $S(0)$, $E(0)$, $I(0)$, and $\rho$.  This signifies the importance of carefully selecting $\sigma_\xi$ in the parameter tracking scheme for this example in order to obtain a useful result.

\end{document}